\RequirePackage{fix-cm}
\documentclass[twocolumn]{svjour3}          
\smartqed  
\usepackage{graphicx}
\usepackage{amssymb}
\usepackage{color}
\usepackage[authoryear]{natbib}
\begin{document}

\title{The Gamma renewal process as an output of the diffusion leaky integrate-and-fire neuronal model
\thanks{This work was supported by the Institute of Physiology RVO:67985823, by the Czech Science Foundation project 15-08066S and University of Torino local grant: ZUCC\_RIC\_LOC\_15\_01.
}
}

\titlerunning{The Gamma renewal process as an output of the diffusion LIF neuronal model}        

\author{Petr Lansky        \and
        Laura Sacerdote \and Cristina Zucca 
}


\institute{P. Lansky \at
              Institute of Physiology, Academy of Sciences of Czech Republic
Vidensk\'{a} 1083, 142 20 Prague 4, Czech Republic \\
              \email{lansky@biomed.cas.cz}           
           \and
           L. Sacerdote and C.Zucca \at
              Department of Mathematics \lq\lq G. Peano\rq\rq, University of Torino, Via Carlo Alberto 10, 10123 Torino, Italy\\
\email{laura.sacerdote@unito.it} \\
\email{cristina.zucca@unito.it}
}

\date{Received: date / Accepted: date}

\maketitle

\begin{abstract}
Statistical properties of spike trains as well as other neurophysiological
data suggest a number of mathematical models of neurons. These models
range from entirely descriptive ones to those deduced from the properties
of the real neurons. 
One of them, the diffusion leaky integrate-and-fire neuronal model, which is based on the Ornstein-Uhlenbeck stochastic process that is restricted by an absorbing barrier, can describe a wide range of neuronal activity in terms of its parameters. These parameters are readily associated with known
physiological mechanisms. The other model is descriptive, Gamma renewal
process, and its parameters only reflect the observed experimental data or
assumed theoretical properties. Both of these commonly used
models are related here. We show under which conditions the Gamma model is
an output from the diffusion Ornstein-Uhlenbeck model. In some cases we can see that the Gamma distribution is unrealistic to be
achieved for the employed parameters of the Ornstein-Uhlenbeck process. 

\end{abstract}

\section{Introduction}
\label{intro}
Numerous models of neuronal spiking activity based on very different
assumptions with different resemblance to reality exist \citep{S}.
This is a natural situation and no one is surprised as the suitability of
a model depends on a purpose for which it has been developed. On the other
hand, it has been always important to point out the bridges, to find
connections among different models, as finally, all of them should stem
out of the same principles. One example of such an effort are the studies
on reduction of the Hodgkin-Huxley model \citep{KAM,KGv,JLG}. For a similar purpose, we recently investigated \citep{RL} the behavior of Stein's neuronal model \citep{St}, which
is based on the leaky integrate-and-fire principle, under the condition
that its input is the output of the model itself.  Our aim here is similar
asking what is the connection between the commonly used Ornstein-Uhlenbeck
(OU) neuronal model and the model of interspike intervals (ISIs) based on
the Gamma renewal process. The choice is not coincidental as both of these
simple models are quite often applied for description of experimental data
as well as for theoretical studies on neuronal coding.

Many references for the OU neuronal model can be given, here are only a
few examples (\citet{RS,I,SSF,LS,VL,Sm}; 
and a recent review \citet{SG}, where
many other references can be found). The model stems out from the OU
stochastic process \citep{UO} which is restricted by
an upper boundary, representing the firing threshold, crossings of which
are identified with generation of spikes. After a spike is initiated, the
memory, including the input, is cleared and the system starts anew.
Therefore, the sequence of ISIs creates a renewal process. Despite the
fact that the OU model coincides with the Langevin equation, it was
originally derived directly from the Stein's neuronal model and thus its
parameters have a clear physiological interpretation \citep{TR,LD}. The OU model has been
generalized in many directions to take into account very different
features of neurons which are not depicted in its basic form. Among these
many variants, important for our purpose, are the models with time-varying
input and time-varying firing threshold. 
The history of time-dependent thresholds in neural models is very long (for recent reviews see \cite{JJR,BML}). However, except the attempts to describe the effect of refractoriness, the dynamical thresholds are aimed at mimicking adaptation in neuronal activity, i.e., a gradual change in the firing rate. Typically, it has been modeled as a decaying single \citep{CPL} or double-exponential \citep{KTS} function.  This is usually accompanied by introducing a correlation structure in a sequence of ISIs (for a review see \cite{AC}). It should be stressed that the time-dependent threshold appearing here is of entirely different nature. Whatever the stimulus is applied at the input, the constant firing rate appears at the output. Simultaneously, the investigated model generates independent and identically distributed ISIs.
While the time-dependent threshold is used to reflect the existence of refractoriness in the behavior of real neurons, the time variable input naturally describes time-varying intensities of the impinging postsynaptic potentials arriving from other neurons in the system \citep{T,SPTTS,PG,Li,B,BCNP,IDL,Th}. Dealing with these models we have
 to
return to the theoretical results on the so called Inverse first-passage-time
problem. 
These results permit us to deduce the shapes of
these functions under the condition that the output of the model is the
Gamma renewal process.

The renewal stochastic process with Gamma distributed intervals between
events is usually called the Gamma (renewal) process (Yannaros, 1988;
Gourevitch and Eggermont, 2007; Farkhooi et al., 2009, Koyama and Kostal,
2014). This model, which we aim to relate to the OU neuronal model, is of
a different nature. It has never been constructed from biological
principles, but has been widely accepted in neuronal context as a good
descriptor of the experimental data and also as a suitable descriptor of
data used in theoretical studies. It appeared immediately when the Poisson
process of ISIs was disregarded as a too simplified description of reality.
The selected references are, similarly to the OU model, only a sample from a much longer list \citep{L,BaL,KA,MA,SB,BeL,SKS,OS}.

As mentioned, our aim is to relate the OU and the Gamma models. More
specifically, we ask under which conditions the OU model generates the
Gamma renewal process as an output. 
The problem was already mentioned in \citep{SZ} but only to illustrate the proposed method. Here the aim is to understand the role of the parameters of the process and of the Gamma distribution.
The relevant properties of both models
are summarized in the first part of the paper. Then, the method to solve
the problem is presented. Finally, the results are illustrated on two
examples and their consequences are shortly discussed.

\section{Ornstein-Uhlenbeck model and Gamma model}

\subsection{The Ornstein-Uhlenbeck model}
The OU stochastic process is a classical model of the membrane potential evolution. It describes the membrane potential $X(t)$ through the one-dimensional process solving the stochastic differential equation

\begin{eqnarray}\label{X}
dX(t)=\left(-\frac{X(t)}{\tau}+\mu\right)dt+\sigma dW(t)
\end{eqnarray}
with initial condition $X(0)=x_0$. Here $\tau>0$ is the membrane time constant, $\mu$ and $\sigma>0$ are two constants that account for the mean and the variability of the input to the neuron and $W(t)$ denotes a standard Wiener process. Further, the model assumes that after each spike the membrane potential is reset to the resetting value $x_0$. In absence of the external input, the membrane potential decays exponentially to the resting potential, which in equation (\ref{X}) is set to zero. The OU process is a continuous Markov process characterized by its transition probability density function (pdf)

\begin{eqnarray}\label{pdf}
f(x,t|x_0)&=&\frac{dP(X(t)<x|X(0)=x_0)}{dx} \\
&=&\frac{1}{\sqrt{ \pi\sigma^2 \tau(1-e^{-2t/\tau})}}\cdot\nonumber\\
&\cdot&\exp \left[ \frac{(x-x_0 e^{-t/\tau}-\mu \tau(1-e^{-t/\tau}))^2}{\sigma^2 \tau(1-e^{-2t/\tau})}\right].\nonumber
\end{eqnarray}
Hence it is Gaussian  and the mean membrane potential is

\begin{eqnarray}\label{EX}
\mathbb{E}\left(X(t)|X(0)=x_0\right)=x_0 e^{-t/\tau}+\mu \tau(1-e^{-t/\tau}),
\end{eqnarray}
and its variance is
\begin{eqnarray}
V\left(X(t)|X(0)=x_0\right)=\frac{\sigma^2 \tau}{2} (1-e^{-2t/\tau}).
\end{eqnarray}

The ISIs generated by the model are identified with the first-passage times (FPTs) $T$ of the process $X(t)$ through a boundary $S$, often taken to be constant
\begin{eqnarray}\label{T}
T=\inf\{t>0:X(t)>S;x_0<S\}.
\end{eqnarray}
Unfortunately the distribution of $T$ is not known in a closed form but its Laplace transform is available, as well as the expressions for the mean $\mathbb{E}(T)$ and the variance $Var(T)$. 
In the presence of the boundary $S$ different dynamics of $X(t)$ arise according to the values of $S$ and $\mu \tau$, the asymptotic value of $\mathbb{E}\left(X(t)|X(0)\right)$. When $\mu \tau<S$, i.e. in the subthreshold regime, the boundary crossings are determined by the noise and the number of spikes in a fixed interval exhibits a Poisson-like distribution. On the contrary, when $\mu \tau>S$ the ISIs are strongly influenced by the input $\mu$ and we speak of suprathreshold regime. 
This division on supra- and sub-threshold regimens is based on the mean behavior of the membrane potential and it evokes the question how it influences the distribution of the FPT.

In some generalizations the model assumes the presence of a time-dependent threshold $S(t)$ or of a time-depending mean input $\mu(t)$. In the latter cases, the process is solution of an equation analogous to (\ref{X}) but with $\mu(t)$ in substitution of $\mu$. In both these cases, the closed forms for the mean and the variance of ISIs analogous to those for (\ref{X}) and (\ref{T}) are no more available. However, suitable numerical methods for determining the FPTs distribution exist as well as reliable simulation techniques. From a mathematical point of view, the case of time-dependent input and that of time-varying threshold are related. In fact, the OU model with time dependent-boundary $S(t)$ and in absence of input, $\mu=0$, (we refer to the case $\mu=0$ because the case $\mu \neq0$ can be obtained from it through a simple transformation), 
\begin{eqnarray}\label{X1}
\left\{
\begin{array}{ll}
dX(t)=\left(-\frac{X(t)}{\tau}\right)dt+\sigma dW(t)\\
X(0)=x_0\\
S(t)
\end{array}
\right.
\end{eqnarray}
can be transformed into the OU model characterized by time-dependent input $\mu(t)$ and constant threshold $\Sigma$
\begin{eqnarray}\label{X2}
\left\{
\begin{array}{ll}
dY(t)=\left(-\frac{Y(t)}{\tau}+\mu(t)\right)dt+\sigma dW(t)\\
Y(0)=x_0-S(0)+\Sigma\\
\Sigma
\end{array}
\right..
\end{eqnarray}
through the space transformation 
\begin{equation}\label{transf}
y=x-S(t)+\Sigma.
\end{equation}
The relationship between the input $\mu(t)$ in (\ref{X2}) and the threshold $S(t)$ in (\ref{X1}) becomes
\begin{equation}\label{mu}
\mu(t)=\frac{\Sigma-S(t)}{\tau}-\frac{dS(t)}{dt}
\end{equation}
which can be integrated to give $S(t)$ explicitly in terms of $\{ \mu(u)\}_{0\leq u\leq t}$ and $S(0)$.

Note that since (\ref{transf}) is a space  transformation, it does not change the FPT distribution of the random variable $T$. 

\subsection{Gamma model of interspike intervals}

A random variable $T$ is Gamma distributed if its pdf is
\begin{equation}
f_T(t)=\frac{\alpha^m}{\Gamma(m)}t^{m-1}e^{-\alpha t}, \quad \quad t\geq 0.
\end{equation}
Here $\alpha>0$ is the rate parameter and $m>0$ is the shape parameter.
Such a random variable is characterized by the following mean, variance and coefficient of variation 
\begin{eqnarray}
E(T)&=\frac{m}{\alpha}\label{ET}\\
Var(T)&=\frac{m}{\alpha^2}\label{VT}\\
CV&=\frac{1}{\sqrt{m}}\label{CV}.
\end{eqnarray}

\begin{figure}[htp]
\centering
\includegraphics[height=15cm]{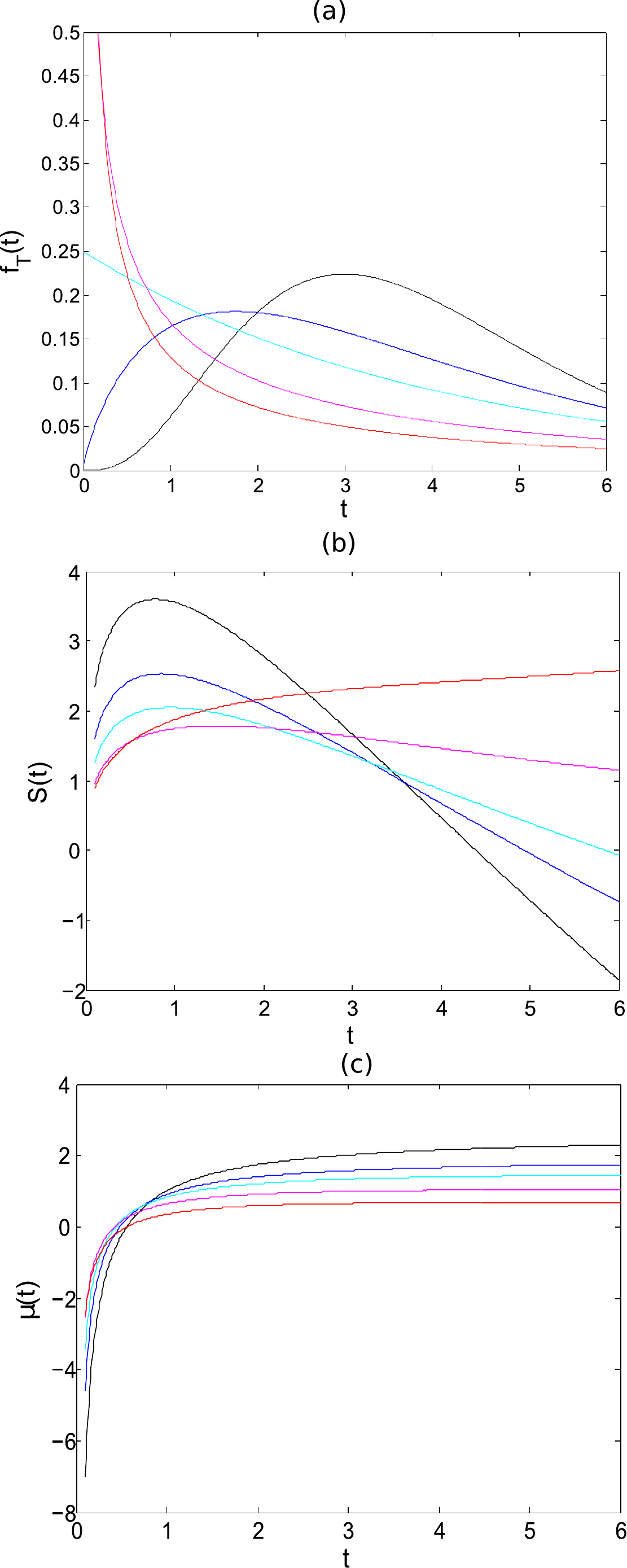}
\caption{Probability densities (a) and evaluated boundaries (b) or mean inputs (c) in the case of Gamma distributed ISIs with mean ISI equal to 4. Different lines correspond to different shapes of the Gamma densities, $CV = 0.5$ (black), $CV=0.75$ (blue), $CV=1$, (cyan), $CV = 1.5$ (magenta), $CV = 2$ (red). The parameters of the OU model are $\tau=10$, $\sigma^2 = 2.5$ and $\Sigma=10$ if the time-variable input is searched for and $\mu=0$ if the time-variable threshold is deduced. 
}
\label{Fig:input}
\end{figure}

The shapes of Gamma pdf for different values of the parameters $\alpha$ and $m$ can be seen in Figure \ref{Fig:input}(a). These shapes strongly change with the value of CV. We recall that $CV=1$ corresponds to ISIs exponentially distributed while when $CV>1$ bursting activity can be modeled and for decreasing CV the activity tends to regularity. 

\section{Results}

\subsection{Theoretical considerations}
We study here, under which conditions the Gamma model of ISIs could be generated by the OU model. 
More specifically we investigate two closely related possibilities:
\begin{enumerate}
\item the membrane potential evolves according to an OU process with $\mu=0$ and we ask if there is a time-dependent threshold $S(t)$ such that the ISIs are Gamma distributed with prescribed parameters;
\item the membrane potential evolves according to an OU process with time-dependent input $\mu(t)$ and spikes are determined by the crossing of a constant threshold $S$, the question remains the same.
\end{enumerate}
The absence of $\mu$ in the first case is only formal due to the transformation of the resting level to zero. Actually, solving one of the problems gives a solution to the other as mentioned, see equation (\ref{transf}).
The case of time-varying input is related to the input restart with a spike. It could arise in a biologically plausible way if the neuron's spike provided input to a second neuron, for instance an inhibitory interneuron, in such a way as to immediately suppress the next spike (for $CV < 1$) or a reciprocally connected excitatory neuron, to accelerate the next spike (for $CV > 1$). Alternatively adaptation currents can produce the effects embodied in the time-varying input.
However, the time-varying threshold does not require any interpretation being a phenomenological quantity and thus we sketch the theoretical method for determining the boundary shape only.
The existence and uniqueness of the solution of the Inverse first-passage-time problem is shown for any regular diffusion process in \citep{CCCS}. Two computational methods to determine the boundary shape
of a Wiener process characterized by an assigned  FPT distribution are presented in \citep{ZS}
while the extension of these methods to the case of an OU process is considered in \citep{SZ}. In that paper the use of the proposed algorithm is illustrated through two examples: the inverse Gaussian distribution and the Gamma distribution. Further related results are in  \citep{SVZ,A}. 

Our task is to determine the time-dependent boundary $S(t)$. To deal with this problem we consider the Fortet integral equation \citep{SG}
\begin{equation}
f(x,t|x_0)=\int_0 ^t  g(S(u),u|x_0)f(x,t|S(u),u)du,
\end{equation}
that relates the transition pdf $f(x,t|x_0)$ of an OU process originated in $x_0$ at time $0$, as given by equation (\ref{pdf}) with the FPT pdf $g(S(u),u|x_0)$ of the process through $S(u)$. This equation holds for any $x \geq S(t)$ and $t>0$. After integrating the Fortet equation with respect to $x$ in $(S(t),\infty)$, we get
\begin{eqnarray}\label{Fortet}
&&1-F(S(t),t|x_0)\\
&&\quad=\int_0 ^t  g(S(u),u|x_0)[1-F(S(t),t|S(u),u)]du \nonumber
\end{eqnarray}
where $F(S(t),t|S(u),u)$ is the transition probability distribution function.
This last equation is a linear Volterra integral equation of the first type where the unknown is the pdf $g(S(t),t|x_0)$, while it is a non-linear Volterra integral equation of the second kind in the unknown $S(t)$. In the following examples we use the algorithm proposed in \citep{SVZ,ZS} to solve the inverse FPT problem and to determine $S(t)$. 
We underline that we fix the Gamma density as a FPT density and we solve the equation (\ref{Fortet}) where the unknown function is the boundary $S(t)$ and $g(S(u),u|x_0)$ is the Gamma density. 
We introduce a time discretization $t_i=ih$ for $i=1,2,\dots$, of step $h>0$ and we discretize the integral equation.
Then, the error of the method concerns the boundary. 
The study of the order of the error on the boundary has been done for a Wiener process in the paper \citep{ZS}. It is proved that the error at each time step $t_n$ is $|S(t_n)-\hat{S}(t_n)|\sim O(h^2)$ where $\hat{S}$ is the approximated boundary obtained by the algorithm. Reproducing the same proof for the OU process, it is possible to prove the same error order also for our process. 
Note that the error does not depend on the parameters, it is due to the right-hand rectangular rule for the discrtization of the integral in (\ref{Fortet}) (cf. \cite{At}).
The algorithm can be applied to every kind of densities, even narrow densities are not a problem. It is sufficient to take a smaller discretization step in order to take into account all the shape of the density. The correct discretization step could be chosen looking at the shape of the FPT density.

We explicitly underline that the algorithm does not require the knowledge of $S(0)$ since it is built with the right-hand
rectangular rule \citep{At}, i.e. the formula does not use the value of the boundary on the l.h.s. of the interval. No condition on $S(0)$ is a numerical advantage because this value is often not known a priori.

The Gamma density can either start from zero, from a constant or from infinity. These different shapes of the Gamma arise in correspondence to different behaviors of the boundary in the origin. 
If the density is null in zero, the corresponding boundary is strictly positive: the density null in zero implies the absence of a probability mass in zero, i.e. no crossing happens for such times. 
On the contrary, a positive mass in the origin requires a boundary starting from zero typically with infinite derivative. In \cite{P} is shown that the only option to get a positive mass for arbitrary small times is to allow the boundary to start together with the process. In order not to have an immediate crossing of all possible trajectories, the boundary should have infinite derivative in zero.
Exponential distribution is often suggested to model ISIs without considering the implication of the choice of a density positive at time zero. 
The behavior of the threshold is interesting from a theoretical point of view, but biologically it carries a limited information as the model is hardly suitable in a close proximity of a previous spike.

The algorithm at time $t_1$ always gives a positive and finite boundary $S(t_1)$. The values of $g(S(t),t|0)$ as $t$ goes to zero may exhibit three different behaviors: in most instances $g(S(0),0|0)=0$, in other cases $g(S(0),0|0)=c>0$ and in the last one $$\lim_{t\rightarrow 0^+}g(S(t),t|0)=\infty.$$
The algorithm is valid if $g(S(0+),0+|0) = 0$ although the more $g(S(0+),0+|0)$ is away from zero, the less accurate the estimation of the boundary through the algorithm becomes in a neighborhood of the origin. Numerically we identify $0^+$ with $t_1$.
Since the algorithm is adaptive, if the discretization step is small enough the error done in a neighborhood of the origin is negligible. 
In Figure \ref{Fig:input}(b) the boundaries corresponding to different Gamma pdf are shown. Since the first steps of the algorithm are not reliable due to the imprecise approximation of the integral in (\ref{Fortet}) in the algorithm, we skip the first interval $[0, 0.1]$. To improve the boundary estimation for small times, we refer to Remark 5.4 in \cite{ZS}.
In Figure \ref{Fig:input}(c) the input functions $\mu(t)$ are plotted. They have been just derived from the boundaries in Figure \ref{Fig:input}(b) by applying (\ref{mu}) and they strongly depend on $dS(t)/dt$.

Since the boundary close to zero cannot be deduced reliably, its
derivative may have a substantial bias here. The algorithm is
self-adaptive, therefore the results become more precise with increasing
time.
 

In Figure \ref{Fig:input}(b) we also note that the thresholds in general decrease as $t$ increases. It may correspond to a weak facilitation of the spiking activity, avoiding the presence of very long ISIs. However, in extreme cases the threshold reaches negative values, which means going below the resting level, and it looks quite unrealistic. The explanation of this feature is straightforward if we take into account Figure \ref{Fig:input}(a) and simultaneously realize what is the input ($\mu = 0$) and the parameter of the underlying OU model ($\tau=10$, $\sigma^2=2.5$). These values would correspond to a strongly subthreshold regimen if a constant threshold ($\Sigma=10$, like for variable $\mu$) is considered and thus the spiking activity would be Poissonian. Under such a scenario the threshold must go in the direction of the mean depolarization ($\mu\cdot \tau=0$) to get Gamma distribution which looks almost Gaussian ($CV=0.5$ and $CV=0.75$). In conclusion, the obtained result of a very negative threshold is an indicator of the unrealism of the hypothesized distribution in the case of employed OU parameters. One cannot obtain in OU model almost regular firing for low signal unless the threshold is substantially modified.

There is a common substantial increase of the threshold after spike generation in Figure \ref{Fig:input}(b) observed mainly for those thresholds which ultimately go to the negative values. Here the same arguments can be presented: while mathematically the model can be forced to produce any shape of Gamma output, biologically it would require a speculative interpretation. 

Note that a boundary becoming negative, despite it can be seen as biologically difficult to interpret, represents no formal problem. The
reason is that there are still numerous trajectories of the OU process below zero and thus below the threshold. Therefore the negative threshold gradually absorbs these trajectories (hyperpolarized below zero) and it creates the tail of the FPT density corresponding to the Gamma distribution.

\subsection{Examples}
We apply the method proposed in the previous section to the Gamma pdfs of
different shapes and show their effect on the shapes of the variable-input
and the variable-threshold. The shape of the Gamma distributions was
varied by changing its CV while its mean was kept constant. The
corresponding parameters can be deduced from equations (\ref{ET}) and (\ref{CV}).
The pdfs of $T$ are plotted for different values of $CV$ in Figure 1(a) and the related boundaries $S(t)$ and the input functions $\mu(t)$, making use of (\ref{EX}), respectively, are plotted in Figure 1(b) and 1(c). The shape of the boundary corresponding to lower values of CV presents a maximum that tends to disappear as CV grows to higher values. Hence, for small values of CV the growth of the boundary eliminates short ISIs. After a certain period of time the thresholds decrease facilitating the attainment of the maxima of the pdf. The Gamma density has no maxima for $CV>1$ and the time-variable threshold tends to become flat. In all cases a complementary behavior is exhibited by the input $\mu(t)$. 

\begin{figure}[htp]
\centering
\includegraphics[height=15cm]{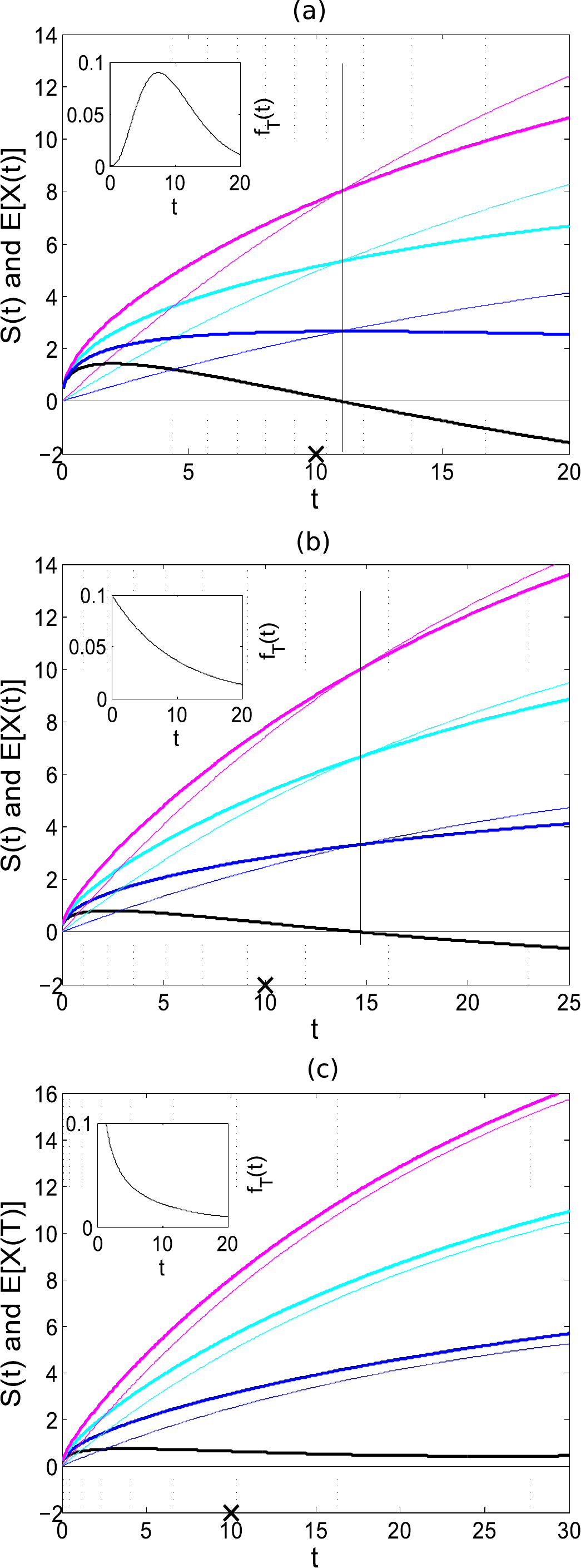}
\caption{The mean membrane potential (\ref{EX}) and the time-variable boundaries corresponding to the Gamma distributed ISIs of three different shapes, $CV = 0.5$ (a), $CV=1$ (b) and $CV=1.5$ (c), the probability densities are on the subplots. The parameters of the OU model are $\tau = 25$, $\sigma^2=0.16$, $x_0 = 0$, the mean ISI is equal to $10$ in all cases represented by a cross on the horizontal axes. Different lines correspond to different values of the mean input: $\mu=0$ (black), $\mu=0.3$ (blue), $\mu=0.6$ (cyan), $\mu=0.9$ (magenta). Thicker lines correspond to the time-variable boundaries. The vertical dotted lines give the ISI distribution quantiles.}
\label{Fig:CV}
\end{figure}

In the second example we investigate not only the behavior of the time-variable firing threshold but also the dynamics of the underlying OU neuronal model. The mean membrane potentials (\ref{EX}) and the boundaries corresponding to the Gamma distributed ISIs with mean ISI equal to $10$ but different values of the mean input $\mu$ and for three different values of $CV$ are shown in Figure 2. The
shapes of the thresholds in Figure \ref{Fig:CV} are concave and initially increasing.
For low input (when $\mu$ is small enough) the curves exhibit a maximum after
which the firing threshold starts to decrease. From a biological view
point such a decrease may be interpreted as reflecting an adaptation
phenomenon. This adaptation is meant in a sense that if there is no spike for a period, then the system becomes more sensitive by decreasing the firing threshold. Initially, the distance between the mean of the membrane
potential and the threshold increases, however, after a certain period, in
all the cases the mean membrane potential crosses the threshold. For fixed
$CV$ this always happens at the same time. This fact can be easily
understood by noting that a change of $\mu$ determines the same shift both
on the mean potential and on the boundary shape,  see (\ref{transf}). This
moment of crossing between the mean potential and the threshold increases
with increasing $CV$. So, practically, for large $CV$ the regimen is again
subthreshold (see Figure \ref{Fig:CV}(c)) for all generated ISIs, whereas for low $CV$
there is a change of the regimen shortly after the mean ISI (see Figure \ref{Fig:CV}(a)
and \ref{Fig:CV}b). This is entirely a new phenomenon if compared with the classical
OU model. Further, we can see that for low $CV$ very short ISIs are rather
improbable due to the presence of an higher threshold for a short time. In general, all the thresholds are initially increasing and that is in contrast with the previously employed time-varying thresholds in the OU neuronal model.

\section{Conclusions}

We presented a method how to modify the OU neuronal model, which is one of the most common models for description of spike generation, to achieve at its output the Gamma renewal process of ISIs. The method is based on the Inverse FPT problem and uses the time-variable firing threshold or time-variable input. While the time-variable input rather lacks a clear biological interpretation, the time-varying firing threshold has been commonly accepted. However, the previous generalizations of the OU model based on introduction of the time-variable threshold always aimed to make the model more realistic \textit{a priori}, whereas here it comes out as a result of a requirement to identify the output of the model with the observed or expected data.

In some parameter cases, the OU process is incapable of generating
gamma-distributed ISIs, unless unrealistic features of the model are
employed. For example, the threshold getting below the resting level is an
indicator of this situation. Interestingly, in other cases, the threshold
corresponding to Gamma distributed ISIs may have a biologically interpretable shape.
It decreases with time and this could be related with neuronal
adaptability. However, this effect does not last over a single ISI and
thus cannot be interpreted as decreasing the firing rate over a spike
train.
Any further interpretation of the threshold behavior would be difficult and surely beyond the scope of this article. On the other hand, it is obvious that it at least partly changes so often applied concept of sub- and supra-threshold firing regimen. Nevertheless, the results of this paper implies that Gamma distributed ISI generated by OU neuronal model with $CV > 1$ are noise driven in contrast to those with $CV \leq 1$ which are driven by both, the signal and the noise.



\end{document}